\documentclass[aps,pre,showpacs,twocolumn]{revtex4}

\usepackage[T1]{fontenc}
\usepackage{amssymb}
\usepackage{amsbsy}
\usepackage{amsmath}
\usepackage{graphicx}
\usepackage{hyperref}    

\begin{document}


\def\beq{\begin{equation}}
\def\eeq{\end{equation}}
\def\llbrace{\left\lbrace}
\def\rrbrace{\right\rbrace}
\def\lbraket{\left[}
\def\rbraket{\right]}

\newcommand{\Tr}{{\rm Tr}} 
\newcommand{\mean}[1]{\langle #1 \rangle}

\newcommand{\ie}{i.e. }
\newcommand{\eg}{e.g. }
\newcommand{\cc}{{\rm c.c.}} 
\newcommand{\hc}{{\rm h.c.}} 

\def\eps{\epsilon}
\def\gam{\gamma} 
\def\phibf{\boldsymbol{\phi}}
\def\varphibf{\boldsymbol{\varphi}}
\def\psibf{\boldsymbol{\psi}}
\def\lamb{\lambda}
\def\sig{\sigma}

\def\half{\frac{1}{2}}

\def\a{{\bf a}}
\def\b{{\bf b}}
\def\e{{\bf e}}
\def\f{{\bf f}}
\def\g{{\bf g}}
\def\h{{\bf h}}
\def\k{{\bf k}}
\def\l{{\bf l}}
\def\m{{\bf m}}
\def\n{{\bf n}} 
\def\p{{\bf p}} 
\def\q{{\bf q}}
\def\r{{\bf r}}
\def\t{{\bf t}}
\def\u{{\bf u}}
\def\v{{\bf v}}
\def\x{{\bf x}}
\def\y{{\bf y}} 
\def\z{{\bf z}} 
\def\A{{\bf A}}
\def\B{{\bf B}}
\def\D{{\bf D}} 
\def\E{{\bf E}} 
\def\F{{\bf F}} 
\def\H{{\bf H}}  
\def\J{{\bf J}}
\def\K{{\bf K}} 

\def\G{{\bf G}}
\def\L{{\bf L}}
\def\M{{\bf M}}  
\def\O{{\bf O}} 
\def\P{{\bf P}} 
\def\Q{{\bf Q}} 
\def\R{{\bf R}}
\def\S{{\bf S}}
\def\nablabf{\boldsymbol{\nabla}}

\def\para{\parallel}

\def\w{\omega}
\def\wn{\omega_n}
\def\wnu{\omega_\nu}
\def\wp{\omega_p} 
\def\dmu{{\partial_\mu}}
\def\dl{{\partial_l}}  
\def\dt{\partial_t} 
\def\tdt{\tilde\partial_t}
\def\dk{\partial_k}
\def\tdk{\tilde\partial_k}
\def\dx{\partial_x}
\def\dy{\partial_y} 
\def\dtau{{\partial_\tau}}  
\def\det{{\rm det}} 
\def\Pf{{\rm Pf}}

\def\intr{\int d^dr}  
\def\dintr{\displaystyle \int d^dr} 
\def\dinttau{\displaystyle \int_0^\beta d\tau}
\def\inttau{\int_0^\beta d\tau}
\def\intx{\int d^{d+1}x} 
\def\inttaur{\int_0^\beta d\tau \int d^dr}
\def\intinf{\int_{-\infty}^\infty}

\def\calA{{\cal A}} 
\def\calC{{\cal C}} 
\def\dt{\partial_t}
\def\calD{{\cal D}}
\def\calF{{\cal F}} 
\def\calG{{\cal G}}
\def\calH{{\cal H}}
\def\calJ{{\cal J}}
\def\calL{{\cal L}} 
\def\calN{{\cal N}}
\def\calO{{\cal O}}
\def\calP{{\cal P}}  
\def\calR{{\cal R}} 
\def\calS{{\cal S}}
\def\calT{{\cal T}}
\def\calU{{\cal U}}
\def\calY{{\cal Y}} 
\def\calZ{{\cal Z}} 

\def\Sign{\Sigma_{\rm n}} 
\def\Sigan{\Sigma_{\rm an}} 
\def\Signt{\tilde\Sigma_{\rm n}} 
\def\Sigant{\tilde\Sigma_{\rm an}} 
\def\Gn{G_{\rm n}}
\def\Gan{G_{\rm an}}
\def\Itildell{\tilde I_{k,ll}}
\def\Itildett{\tilde I_{k,tt}}
\def\Jtildellll{\tilde J_{k;ll,ll}}
\def\Jtildetttt{\tilde J_{k;tt,tt}}
\def\Jtildeltlt{\tilde J_{k;lt,lt}}
\def\Jtildelltt{\tilde J_{k;ll,tt}}
\def\Jtildettll{\tilde J_{k;tt,ll}}
\def\Jtildelllt{\tilde J_{k;ll,lt}}
\def\Jtildeltll{\tilde J_{k;lt,ll}}
\def\Jtildettlt{\tilde J_{k;tt,lt}}
\def\Jtildelttt{\tilde J_{k;lt,tt}}

\title{Infrared behavior of interacting bosons at zero temperature}
  
\author{N. Dupuis}
\author{A. Ran\c{c}on}
\affiliation{Laboratoire de Physique Th\'eorique de la Mati\`ere Condens\'ee, 
CNRS - UMR 7600, \\ Universit\'e Pierre et Marie Curie, 4 Place Jussieu, 
75252 Paris Cedex 05, France}

\date{28 October 2010}

\begin{abstract}
We review the infrared behavior of interacting bosons at zero temperature. After a brief discussion of the Bogoliubov approximation and the breakdown of perturbation theory due to infrared divergences, we present two approaches that are free of infrared divergences -- Popov's hydrodynamic theory and the non-perturbative renormalization group -- and allow us to obtain the exact infrared behavior of the correlation functions. We also point out the connection between the infrared behavior in the superfluid phase and the critical behavior at the superfluid--Mott-insulator transition in the Bose-Hubbard model. 
\end{abstract}


\maketitle

\section{Introduction}

Many of the predictions of the Bogoliubov theory of superfluidity~\cite{Bogoliubov47} have been confirmed experimentally, in particular in ultracold atomic gases~\cite{Dalfovo99,Leggett01}. Nevertheless a clear understanding of the infrared behavior of interacting bosons at zero temperature has remained a challenging theoretical issue for a long time. Early attempts to go beyond the Bogoliubov theory have revealed a singular perturbation theory plagued by infrared divergences due to the presence of the Bose-Einstein condensate and the Goldstone mode~\cite{Hugenholtz59,Gavoret64,Beliaev58a,Beliaev58b}. In the 1970s, Nepomnyashchii and Nepomnyashchii proved that the anomalous self-energy vanishes at zero frequency and momentum in dimension $d\leq 3$~\cite{Nepomnyashchii75}. This exact result shows that the Bogoliubov approximation, where the linear spectrum and the superfluidity rely on a finite value of the anomalous self-energy, breaks down at low energy. As realized latter on~\cite{Nepomnyashchii78,Nepomnyashchii83}, the singular perturbation theory is a direct consequence of the coupling between transverse and longitudinal fluctuations and reflects the divergence of the longitudinal susceptibility -- a general phenomenon in systems with a continuous broken symmetry~\cite{Patasinskij73}. 

In this paper, we review the infrared behavior of interacting bosons. A more detailed discussion together with a comparison to the classical O($N$) model can be found in Ref.~\cite{Dupuis10}. In Sec.~\ref{sec_bosons_pt}, we briefly review the Bogoliubov theory and the appearance of infrared divergences in perturbation theory. We introduce the Ginzburg momentum scale $p_G$ signaling the breakdown of the Bogoliubov approximation. In Sec.~\ref{sec_hydro}, we discuss Popov's hydrodynamic approach based on a phase-density representation of the boson field~\cite{Popov72,Popov79,Popov_book_2}. This approach allows one to derive the order parameter correlation function without encountering infrared divergences. The non-perturbative renormalization group (NPRG) provides another approach free of infrared divergences which yields the exact infrared behavior of the normal and anomalous single-particle propagators (Sec.~\ref{sec_bosons_nprg})~\cite{Castellani97,Pistolesi04,Wetterich08,Floerchinger08,Dupuis07,Dupuis09a,Dupuis09b,Sinner09,Sinner10}. In the last section, we discuss the critical behavior at the superfluid--Mott-insulator transition in the Bose-Hubbard model and its connection to the infrared behavior in the superfluid phase.

\section{Perturbation theory and breakdown of the Bogoliubov approximation}
\label{sec_bosons_pt}

We consider interacting bosons at zero temperature with the (Euclidean) action
\begin{equation}
S = \int dx \left[ \psi^*\left(\dtau-\mu - \frac{\nablabf^2}{2m}
  \right) \psi + \frac{g}{2} (\psi^*\psi)^2 \right] ,
\label{action}
\end{equation}
where $\psi(x)$ is a bosonic (complex) field, $x=(\r,\tau)$, and $\int dx=\inttau \int d^dr$. $\tau\in [0,\beta]$ is an imaginary time, $\beta\to\infty$ the inverse temperature, and $\mu$ denotes the
chemical potential. The interaction is assumed to be local in space and the
model is regularized by a momentum cutoff $\Lambda$. We consider a space dimension $d>1$.

Introducing the two-component field 
\begin{equation}
\Psi(p) = \left( \begin{array}{c} \psi(p) \\ \psi^*(-p)  \end{array} \right) , \quad 
\Psi^\dagger(p) = \bigl( \psi^*(p), \psi(-p) \bigr) 
\end{equation}
(with $p=(\p,i\w)$ and $\w$ a Matsubara frequency), the one-particle (connected) propagator becomes a $2\times 2$ matrix whose inverse in Fourier space is given by
\begin{equation}
\left( 
\begin{array}{cc} i\w + \mu -\eps_\p -\Sign(p) & - \Sigan(p) \\
 -\Sigan^*(p) & -i\w + \mu -\eps_\p -\Sign(-p)
\end{array}
\right) ,
\label{propa}
\end{equation}
where $\Sign$ and $\Sigan$ are the normal and anomalous self-energies, respectively, and $\eps_\p=\p^2/2m$. If we choose the order parameter $\mean{\psi(x)}=\sqrt{n_0}$ to be real (with $n_0$ the condensate density), then the anomalous self-energy $\Sigan(p)$ is real.

\subsection{Bogoliubov approximation} 
\label{subsec_bog}

The Bogoliubov approximation is a Gaussian fluctuation theory about the saddle point solution $\psi(x)=\sqrt{n_0}=\sqrt{\mu/g}$. It is equivalent to a zero-loop calculation of the self-energies~\cite{AGD_book,Fetter_book},
\begin{equation}
\Sign^{(0)}(p) = 2gn_0 , \quad
\Sigan^{(0)}(p) = gn_0 .
\end{equation}
This yields the (connected) propagators 
\beq
\begin{split}
G_{\rm n}^{(0)}(p) &= -\mean{\psi(p)\psi^*(p)}_c = \frac{-i\w-\eps_\p-gn_0}{\w^2+E_\p^2} ,\\ 
G_{\rm an}^{(0)}(p) &= -\mean{\psi(p)\psi(-p)}_c = \frac{gn_0}{\w^2+E_\p^2} ,
\end{split}
\eeq
where $E_\p=[\eps_\p(\eps_\p+2gn_0)]^{1/2}$ is the Bogoliubov quasi-particle excitation energy. When $|\p|$ is larger than the healing momentum $p_c=(2gmn_0)^{1/2}$, the spectrum $E_\p\simeq \eps_\p+gn_0$ is particle-like, whereas it becomes sound-like for $|\p|\ll p_c$ with a velocity $c=\sqrt{gn_0/m}$. In the weak-coupling limit, $n_0\simeq \bar n$ ($\bar n$ is the mean boson density) and $p_c$ can equivalently be defined as $p_c=(2gm\bar n)^{1/2}$. 

It is convenient to write the boson field 
\beq
\psi(x) = \frac{1}{\sqrt{2}} [\psi_1(x) + i\psi_2(x)] 
\label{psi12}
\eeq
in terms of two real fields $\psi_1$ and $\psi_2$, which allows us to distinguish between longitudinal ($\psi_1$) and transverse ($\psi_2$) fluctuations. In the hydrodynamic regime $|\p|\ll p_c$,
\beq
\begin{split}
G_{11}^{(0)}(p) &= \frac{\eps_\p}{\w^2+c^2 \p^2} , \\ 
G_{22}^{(0)}(p) &= \frac{2gn_0}{\w^2+c^2 \p^2} , \\ 
G_{12}^{(0)}(p) &= -\frac{\w}{\w^2+c^2 \p^2} ,
\end{split}
\label{Ghydro}
\eeq 
where $G_{ij}(\p)=\mean{\psi_i(\p)\psi_j(-\p)}_c$. In the Bogoliubov approximation, the occurrence of a linear spectrum at low energy (which implies superfluidity according to Landau's criterion) is due to $\Sigan(p=0)$ being nonzero. 

\subsection{Infrared divergences and the Ginzburg scale}
\label{subsec_bosons_ir}

\begin{figure}
\centerline{\includegraphics[height=1.8cm]{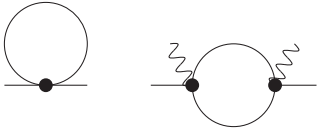}}
\caption{One-loop correction $\Sigma^{(1)}$ to the self-energy. The dots represent the bare interaction, the zigzag lines the order parameter $\sqrt{n_0}$, and the solid lines the connected propagator $G^{(0)}$.}
\label{fig_self_1loop} 
\end{figure}

Let us now consider the lowest-order (one-loop) correction $\Sigma^{(1)}$ to the Bogoliubov result $\Sigma^{(0)}$. For $d\leq 3$, the second diagram of Fig.~\ref{fig_self_1loop} gives a divergent contribution when the two internal lines correspond to transverse fluctuations, which indicates a breakdown of perturbation theory and therefore the Bogoliubov approximation. Retaining only the divergent part, we obtain 
\beq
\Sign^{(1)}(p) \simeq \Sigan^{(1)}(p) \simeq - \half g^2 n_0 \int_q G_{22}^{(0)}(q) G_{22}^{(0)}(p+q) ,
\label{sigma3}
\eeq
where we use the notation $\int_q=\intinf\frac{d\w'}{2\pi} \int_\q$. For small $p$, the main contribution to the $q$ integral in (\ref{sigma3}) comes from momenta $|\q|\lesssim p_c$ and frequencies $|\w'|\lesssim cp_c$. We can then use (\ref{Ghydro}) and obtain 
\beq
\Sign^{(1)}(p) \simeq \Sigan^{(1)}(p) \simeq - 2 \frac{g^4 n_0^3}{c^3} A_{d+1} \left(\p^2+\frac{\w^2}{c^2}\right)^{(d-3)/2} 
\label{sigma4a}
\eeq
if $d<3$ and 
\beq
\Sign^{(1)}(p) \simeq \Sigan^{(1)}(p) \simeq - \frac{g^4 n_0^3}{c^3} A_{4} \ln\left(\frac{p^2_c}{\p^2+\w^2/c^2}\right)  
\label{sigma4b}
\eeq
if $d=3$, where 
\beq
A_d = \llbrace 
\begin{array}{lcc}
- \frac{2^{1-d}\pi^{1-d/2}}{\sin(\pi d/2)} \frac{\Gamma(d/2)}{\Gamma(d-1)} & \mbox{if} & d<4 ,\\ 
\frac{1}{8\pi^2} & \mbox{if} & d=4 .
\end{array}
\right. 
\eeq
We can estimate the characteristic (Ginzburg) momentum scale $p_G$ below which the Bogoliubov approximation breaks down from the condition $|\Sign^{(1)}(p)| \sim \Sign^{(0)}(p)$ or  $|\Sigan^{(1)}(p)| \sim \Sigan^{(0)}(p)$ for $|\p|=p_G$ and $|\w|=cp_G$,
\begin{equation}
p_G \sim \left\lbrace 
\begin{array}{lcc}
(A_{d+1} gm p_c)^{1/(3-d)} & \mbox{if} & d<3 , \\
p_c \exp\left( - \frac{1}{A_4 gmp_c}\right) & \mbox{if} & d=3 .
\end{array}
\right.
\label{pG_est}
\end{equation}
This result can be rewritten as
\begin{equation}
p_G \sim \left\lbrace 
\begin{array}{lcc}
p_c (A_{d+1} \tilde g^{d/2})^{1/(3-d)} & \mbox{if} & d<3 , \\
p_c \exp\left( - \frac{1}{A_4 \sqrt{2} \tilde g^{3/2}}\right) & \mbox{if} & d=3 ,
\end{array}
\right.
\end{equation}
where 
\beq
\tilde g = gm \bar n^{1-2/d} \sim \left(\frac{p_c}{\bar n^{1/d}}\right)^2
\eeq
is the dimensionless coupling constant obtained by comparing the mean interaction energy per particle $g\bar n$ to the typical kinetic energy $1/m\bar r^2$ where $\bar r\sim \bar n^{-1/d}$ is the mean distance between particles~\cite{Petrov04}. A superfluid is weakly correlated if $\tilde g\ll 1$, \ie $p_G\ll p_c\ll \bar n^{1/d}$ (the characteristic momentum scale $\bar n^{1/d}$ does however not play any role in the weak-coupling limit)~\cite{Capogrosso10}. In this case, the Bogoliubov theory applies to a large part of the spectrum where the dispersion is linear (\ie $|\p|\lesssim p_c$) and breaks down only at very small momenta  $|\p|\lesssim p_G\ll p_c$. When the dimensionless coupling $\tilde g$ becomes of order unity, the three characteristic momentum scales $p_G\sim p_c\sim \bar n^{1/d}$ become of the same order. The momentum range $[p_G,p_c]$ where the linear spectrum can be described by the Bogoliubov theory is then suppressed. We expect the strong-coupling regime $\tilde g\gg 1$ to be governed by a single characteristic momentum scale, namely $\bar n^{1/d}$.

\subsection{Vanishing of the anomalous self-energy} 
\label{subsec_bosons_sigmaexact}

The exact values of $\Sign(p=0)$ and $\Sigan(p=0)$ can be obtained using the U(1) symmetry of the action, \ie the invariance under the field transformation $\psi(x)\to e^{i\theta}\psi(x)$ and $\psi^*(x)\to e^{-i\theta}\psi^*(x)$. On the one hand, the self-energies satisfy the Hugenholtz-Pines theorem~\cite{Hugenholtz59}, 
\beq
\Sign(p=0)-\Sigan(p=0)=\mu . 
\label{sigma6a}
\eeq
On the other hand, the anomalous self-energy vanishes,
\beq
\Sigan(p=0) = 0 .
\label{sigma6b}
\eeq 
The last result was first proven by Nepomnyashchii and Nepomnyashchii~\cite{Nepomnyashchii75,Sinner10,Dupuis10}. It shows that the Bogoliubov theory, where the linear spectrum and the superfluidity rely on a finite value of the anomalous self-energy, breaks down at low energy in agreement with the conclusions drawn from perturbation theory (Sec.~\ref{subsec_bosons_ir}).

\section{Popov's hydrodynamic theory}
\label{sec_hydro} 

It was realized by Popov that the phase-density representation of the boson field $\psi=\sqrt{n}e^{i\theta}$ leads to a theory free of infrared divergences~\cite{Popov72,Popov_book_2}. In this section, we show how this allows us to obtain the infrared behavior of the propagators $G_{\rm n}(p)$ and $G_{\rm an}(p)$ without encountering infrared divergences~\cite{Popov79}. 

\subsection{Hydrodynamic action} 

In terms of the density and phase fields, the action reads
\beq
S[n,\theta] = \int dx \left[ in \dot\theta + \frac{n}{2m}(\nablabf\theta)^2 + \frac{(\nablabf n)^2}{8mn} - \mu n + \frac{g}{2} n^2 \right] .
\label{action6}
\eeq
At the saddle-point level, $n(x)=\bar n=\mu/g$. Expanding the action to second order in $\delta n=n-\bar n$, $\dot\theta$ and $\nablabf\theta$, we obtain 
\beq
S[\delta n,\theta] = \int dx \left[ i\delta n \dot\theta + \frac{\bar n}{2m}(\nablabf\theta)^2 + \frac{(\nablabf n)^2}{8m\bar n} + \frac{g}{2} (\delta n)^2 \right] .
\label{actionh} 
\eeq
The higher-order terms can be taken into account within perturbation theory and only lead to finite corrections of the coefficients of the hydrodynamic action (\ref{actionh})~\cite{Popov_book_2}.

We deduce the correlation functions of the hydrodynamic variables, 
\beq
\begin{split}
G_{nn}(p) &= \mean{\delta n(p)\delta n(-p)} = \frac{\bar n}{m} \frac{\p^2}{\w^2+E_\p^2} , \\ 
G_{n\theta}(p) &= \mean{\delta n(p)\theta(-p)} = -\frac{\w}{\w^2+E_\p^2} , \\
G_{\theta\theta}(p) &= \mean{\theta(p)\theta(-p)} = \frac{\frac{\p^2}{4m\bar n}+g}{\w^2+E_\p^2} ,
\end{split}
\eeq
where $E_\p$ is the Bogoliubov excitation energy defined in Sec.~\ref{subsec_bog}. In the hydrodynamic regime $|\p|\ll p_c=\sqrt{2gm\bar n}$, 
\beq
\begin{split}
G_{nn}(p) &=  \frac{\bar n}{m} \frac{\p^2}{\w^2+c^2\p^2} , \\ 
G_{n\theta}(p) &= -\frac{\w}{\w^2+c^2\p^2} , \\
G_{\theta\theta}(p) &= \frac{mc^2}{\bar n}\frac{1}{\w^2+c^2\p^2} ,
\end{split}
\label{Gc}
\eeq
where $c=\sqrt{g\bar n/m}$ is the Bogoliubov sound mode velocity. It can be shown that Eqs.~(\ref{Gc}) are exact in the hydrodynamic limit $|\p|,|\w|/c\ll p_c$ provided that $c$ is the exact sound mode velocity and $\bar n$ the actual mean density (which may differ from $\mu/g$)~\cite{Popov_book_2,Dupuis10}.

\subsection{Normal and anomalous propagators}
\label{subsec_bosons_gnan}

To compute the propagator of the $\psi$ field, we write 
\beq
\psi(x) = \sqrt{n_0+\delta n(x)}e^{i\theta(x)}, 
\eeq
where $n_0=|\mean{\psi(x)}|^2=|\mean{\sqrt{n(x)}e^{i\theta(x)}}|^2$ is the condensate density. For a weakly-interacting superfluid, $n_0\simeq \bar n$, and we expect the fluctuations $\delta n$ to be small. Let us assume that the superfluid order parameter $\mean{\psi(x)}=\sqrt{n_0}$ is real. Transverse and longitudinal fluctuations are then expressed as
\beq
\begin{split}
\delta\psi_2 &= \sqrt{2n_0}\theta + \cdots \\ 
\delta\psi_1 &= \frac{\delta n}{\sqrt{2n_0}} - \sqrt{\frac{n_0}{2}} \theta^2 + \cdots 
\end{split}
\eeq
where the ellipses stand for subleading contributions to the low-energy behavior of the correlation functions. For the transverse propagator, we obtain 
\beq
G_{22}(p) \simeq 2n_0 G_{\theta\theta}(p) = \frac{2n_0mc^2}{\bar n}\frac{1}{\w^2+c^2\p^2}
\label{G22}
\eeq
to leading order in the hydrodynamic regime, while
\beq
G_{12}(p) \simeq G_{n\theta}(p) = - \frac{\w}{\w^2+c^2\p^2} . 
\eeq
The longitudinal propagator is given by 
\begin{align}
G_{11}(x) &= \frac{1}{2n_0} G_{nn}(x) + \frac{n_0}{2} \mean{\theta(x)^2\theta(0)^2}_c \nonumber \\ 
&= \frac{1}{2n_0} G_{nn}(x) + n_0 G_{\theta\theta}(x)^2 , 
\end{align} 
where the second line is obtained using Wick's theorem. In Fourier space, 
\begin{align}
G_{11}(p) ={}& \frac{\bar n}{2mn_0}\frac{\p^2}{\w^2+c^2\p^2} + n_0 G_{\theta\theta}\star G_{\theta\theta}(p) ,
\label{G11}
\end{align} 
where
\beq 
G_{\theta\theta}\star G_{\theta\theta}(p) = \int_q G_{\theta\theta}(q)G_{\theta\theta}(p+q) .
\label{GG}
\eeq
The dominant contribution to the integral in~(\ref{GG}) comes from momenta $|\q|\lesssim p_c$ and frequencies $|\w'|/c\lesssim p_c$, \ie
\begin{align}
\lefteqn{ G_{\theta\theta}\star G_{\theta\theta}(p)} \hspace{0.5cm}  & \nonumber \\ &= \llbrace 
\begin{array}{lcc}
 A_{d+1} c\left(\frac{m}{\bar n}\right)^2 \left(\p^2+\frac{\w^2}{c^2}\right)^{(d-3)/2} & \mbox{if} & d<3 , \\ 
\frac{A_4}{2} c \left(\frac{m}{\bar n}\right)^2 \ln \left(\frac{p_c^2}{\p^2+\frac{\w^2}{c^2}}\right)  & \mbox{if} & d=3 .
\end{array}
\right. 
\end{align}
By comparing the two terms in the rhs of (\ref{G11}) with $|\p|=p_G$ and $|\w|=cp_G$, we recover the Ginzburg scale (\ref{pG_est}). For $|\p|,|\w|/c\gg p_G$, the last term in the rhs of (\ref{G11}) can be neglected and we reproduce the result of the Bogoliubov theory (noting that $\bar n\simeq n_0$), while for $|\p|,|\w|/c\ll p_G$, $G_{11}(p)\sim 1/(\w^2+c^2\p^2)^{(3-d)/2}$ is dominated by phase fluctuations (Goldstone regime). The longitudinal susceptibility $G_{11}(\p,i\w=0)\sim 1/|\p|^{3-d}$ for $\p\to 0$ in contrast to the Bogoliubov approximation $G_{11}(\p,i\w=0)=1/2mc^2$. 

From these results, we deduce the hydrodynamic behavior of the normal propagator~\cite{note11},
\begin{align}
G_{\rm n}(p) ={}& - \half \left[ G_{11}(p) -2iG_{12}(p) + G_{22}(p)\right] \nonumber \\ 
={}& - \frac{n_0mc^2}{\bar n} \frac{1}{\w^2+c^2\p^2} \nonumber \\ &
- \frac{i\w}{\w^2+c^2\p^2} - \half G_{11}(p) ,
\label{Gn}
\end{align}
as well as that of the anomalous propagator, 
\begin{align}
G_{\rm an}(p) ={}& - \half \left[ G_{11}(p) - G_{22}(p)\right] \nonumber \\  
&= \frac{n_0mc^2}{\bar n} \frac{1}{\w^2+c^2\p^2} - \half G_{11}(p) , 
\label{Gan}
\end{align}
where $G_{11}(p)$ is given by (\ref{G11}). The leading-order terms in (\ref{Gn}) and (\ref{Gan}) agree with the results of Gavoret and Nozi\`eres~\cite{Gavoret64} and are exact (see next section). The contribution of the diverging longitudinal correlation function was first identified by
Nepomnyashchii and Nepomnyashchii~\cite{Nepomnyashchii78} and later in Refs.~\cite{Popov79,Weichman88,Giorgini92,Castellani97,Pistolesi04}.

\subsection{Normal and anomalous self-energies}
\label{subsec_bosons_self}

To compute the self-energies $\Sign(p)$ and $\Sigan(p)$, we use the relations
\beq
\begin{split}
\Sign(p) &= G_0^{-1}(p) - \frac{\Gn(-p)}{\Gn(p)\Gn(-p)-\Gan(p)^2} , \\
\Sigan(p) &= \frac{\Gan(p)}{\Gn(p)\Gn(-p)-\Gan(p)^2} .
\end{split}
\label{self}
\eeq
This yields~\cite{Popov79,Dupuis10} 
\begin{align}
\Sigan(p) &= \Sign(p) - G_0^{-1}(p) \nonumber \\ &= \llbrace
\begin{array}{lcc} 
\frac{\bar n^2}{2A_{d+1}c^{4-d}n_0 m^2} (\w^2+c^2\p^2)^{(3-d)/2} & \mbox{if} & d<3 , \\ 
\frac{\bar n^2}{A_4 c n_0 m^2} \left[\ln \left(\frac{\w^2+c^2\p^2}{c^2 p_c^2}\right)\right]^{-1} & \mbox{if} & d=3 ,
\end{array}
\right. 
\label{sigma5}
\end{align} 
in the infrared limit $|\p|,|\w|/c\ll p_G$, where $G_0^{-1}(p)=i\w-\eps_\p+\mu$. Equations~(\ref{sigma5}) agree with the exact results (\ref{sigma6a},\ref{sigma6b}) and show that $\Sign(p)$ and $\Sigan(p)$ are dominated by non-analytic terms for $p\to 0$. This non-analyticity reflects the singular behavior of the longitudinal correlation function 
\beq
G_{11}(p) \simeq \frac{1}{2\Sigan(p)} 
\eeq
in the low-energy limit.

\section{The non-perturbative RG} 
\label{sec_bosons_nprg} 

The NPRG provides another way to circumvent the difficulties of perturbation theory and derive the correlation functions in the low-energy limit~\cite{Castellani97,Pistolesi04,Dupuis07,Dupuis09a,Dupuis09b,Wetterich08,Floerchinger08,Sinner09,Sinner10}. The strategy of the NPRG is to build a family of theories indexed by a momentum scale $k$ such that fluctuations are smoothly taken into account as $k$ is lowered from the microscopic scale $\Lambda$ down to 0~\cite{Berges02,Delamotte07}. This is achieved by adding to the action (\ref{action}) an infrared regulator term
\beq
\Delta S_k[\psi] = \half \sum_{p,i} \psi_i(-p) R_k(p) \psi_i(p) ,
\label{irreg}
\eeq
where $\psi_1$ and $\psi_2$ are the two real fields introduced in Sec.~\ref{subsec_bog} [Eq.~(\ref{psi12})]. The main quantity of interest is the so-called average effective action
\beq
\Gamma_k[\phi] = - \ln Z_k[J] + \sum_{p,i} J_i(-p)\phi_i(p) - \Delta S_k[\phi] ,
\eeq
defined as a modified Legendre transform of $-\ln Z_k[J]$ which includes the subtraction of $\Delta S_k[\phi]$. $J_i$ denotes an external source that couples linearly to the boson field $\psi_i$ and $\phi_i(x) = \mean{\psi_i(x)}$ is the superfluid order parameter. The cutoff function $R_k$ is chosen such that at the microscopic scale $\Lambda$ it suppresses all fluctuations, so that the mean-field approximation $\Gamma_\Lambda[\phi]=S[\phi]$ becomes exact. The effective action of the original model (\ref{action}) is given by $\Gamma_{k=0}$ provided that $R_{k=0}$ vanishes. For a generic value of $k$, the cutoff function $R_k(p)$ suppresses fluctuations with momentum $|\p|\lesssim k$ and frequency $|\w|\lesssim ck$ but leaves those with $|\p|,|\w|/c\gtrsim k$ unaffected ($c\equiv c_k$ is the velocity of the Goldstone mode). The dependence of the average effective action on $k$ is given by Wetterich's equation~\cite{Wetterich93} 
\beq
\dt \Gamma_k[\phi] = \half \Tr\llbrace \dot R_k\left(\Gamma^{(2)}_k[\phi] + R_k\right)^{-1} \rrbrace ,
\label{rgeq}
\eeq
where $t=\ln(k/\Lambda)$ and $\dot R_k=\dt R_k$. $\Gamma^{(2)}_k[\phi]$ denotes the second-order functional derivative of $\Gamma_k[\phi]$. In Fourier space, the trace involves a sum over momenta and frequencies as well as the internal index of the $\phi$ field. We choose the cutoff function~\cite{Dupuis09b}
\beq
R_k(p) = \frac{Z_{A,k}}{2m} \left(\p^2+\frac{\w^2}{c_0^2} \right) r\left( \frac{\p^2}{k^2}+\frac{\w^2}{k^2c_0^2} \right) , 
\label{regdef}
\eeq
where $r(Y)=(e^Y-1)^{-1}$. The $k$-dependent variable $Z_{A,k}$ is defined below. A natural choice for the velocity $c_0$ would be the actual ($k$-dependent) velocity $c_k$ of the Goldstone mode. In the weak coupling limit, however, the Goldstone mode velocity renormalizes only weakly and is well approximated by the $k$-independent value $c_0=\sqrt{g\bar n/m}$.

\subsection{Derivative expansion and infrared behavior} 
\label{subsec_bosons_de} 

Because of the regulator term $\Delta S_k$, the vertices $\Gamma^{(n)}_{k,i_1\cdots i_n}(p_1,\cdots,p_n)$ are smooth functions of momenta and frequencies and can be expanded in powers of $\p_i^2/k^2$ and $\w_i^2/c^2k^2$. Thus if we are interested only in the low-energy properties, we can use a derivative expansion of the average effective action~\cite{Berges02,Delamotte07}. In the following we consider the Ansatz 
\begin{align}
\Gamma_k[\phi^*,\phi] = \int dx\Bigl[ & \phi^*\Bigl(Z_{C,k}\dtau - V_{A,k}\partial^2_\tau - \frac{Z_{A,k}}{2m} \nablabf^2 \Bigr) \phi \nonumber \\ & + \frac{\lamb_k}{2} (n-n_{0,k})^2 \Bigr] ,
\label{effaction} 
\end{align} 
where $n=|\phi|^2=\half(\phi_1^2+\phi_2^2)$. $n_{0,k}$ denotes the condensate density in the equilibrium state. We have introduced a second-order time derivative term. Although not present in the initial average effective action $\Gamma_\Lambda$, we shall see that this term plays a crucial role when $d\leq 3$~\cite{Wetterich08,Dupuis07}.

In a broken U(1) symmetry state with order parameter $\phi_1=\sqrt{2n_0}$, $\phi_2=0$, the two-point vertex is given by 
\beq
\begin{split}
\Gamma^{(2)}_{k,11}(p) &= V_{A,k}\w^2+Z_{A,k}\eps_\p + 2\lamb_{k}n_{0,k}  , \\ 
\Gamma^{(2)}_{k,22}(p) &= V_{A,k}\w^2+Z_{A,k}\eps_\p , \\ 
\Gamma^{(2)}_{k,12}(p) &= Z_{C,k}\w .
\end{split}
\label{gamma2}
\eeq
Since these expressions are obtained from a derivative expansion of the average effective action, they are valid only in the limit $|\p|,|\w|/c\ll k$. In practice however, one can retrieve the $p$ dependence of $\Gamma^{(2)}_{k=0}(p)$ at finite $p$ by stopping the RG flow at $k\sim \sqrt{\p^2+\w^2/c^2}$~\cite{Dupuis09b}. From (\ref{gamma2}) we deduce the derivative expansion of the normal and anomalous self-energies, 
\beq
\begin{split}
\Sigma_{k,{\rm n}}(p) ={}& \mu + V_{A,k}\w^2 + (1- Z_{C,k})i\w \nonumber \\ & 
- (1-Z_{A,k})\eps_\p + \lamb_k n_{0,k} , \\ 
\Sigma_{k,{\rm an}}(p) ={}& \lamb_{k}n_{0,k} . 
\end{split}
\eeq
At the initial stage of the flow, $Z_{A,\Lambda}=Z_{C,\Lambda}=1$, $V_{A,\Lambda}=0$, $\lamb_\Lambda=g$   and $n_{0,\Lambda}=\mu/g$, which reproduces the results of the Bogoliubov approximation. 

Since the anomalous self-energy $\Sigma_{k=0,{\rm an}}(p)\sim (\w^2+c^2\p^2)^{(3-d)/2}$ is singular for $|\p|,|\w|/c\ll p_G$ and $d\leq 3$ (see Sec.~\ref{sec_hydro}), we expect $\Sigma_{k,{\rm an}}(p=0)\sim k^{3-d}$ for $k\ll p_G$ (given the equivalence between $k$ and $\sqrt{\p^2+\w^2/c^2}$), \ie 
\beq
\lamb_k \sim k^{3-d} 
\label{lambk1}
\eeq
($n_{0,k=0}$ is finite in the superfluid phase). The hypothesis (\ref{lambk1}) is sufficient, when combined to Galilean and gauge invariances, to obtain the exact infrared behavior of the normal and anomalous propagators. In the domain of validity of the derivative expansion, $|\p|^2,|\w|^2/c^2\ll k^2\ll k^{3-d}$ for $k\to 0$, one obtains from (\ref{gamma2})
\begin{equation}
\begin{split}
G_{k,11}(p) &= \frac{1}{2\lambda_kn_{0,k}} , \\ 
G_{k,22}(p) &= \frac{1}{V_{A,k}} \frac{1}{\w^2+c^2_k\p^2} , \\
G_{k,12}(p) &= -\frac{Z_{C,k}}{2\lambda_k n_{0,k} V_{A,k}} \frac{\w}{\w^2+c^2_k\p^2} ,
\end{split}
\label{de2}
\end{equation}
where
\begin{equation}
c_k = \left( \frac{Z_{A,k}/2m}{V_{A,k}+Z_{C,k}^2/2\lambda_k n_{0,k}} \right)^{1/2} 
\label{cdef} 
\end{equation}
is the velocity of the Goldstone mode. From (\ref{lambk1}) and (\ref{de2}), we recover the divergence of the longitudinal susceptibility if we identify $k$ with $\sqrt{\p^2+\w^2/c^2}$.

The parameters $Z_{A,k}$, $Z_{C,k}$ and $V_{A,k}$ can be related to thermodynamic quantities using Ward identities~\cite{Gavoret64,Huang64,Pistolesi04,Dupuis09b},
\begin{equation}
\begin{split}
n_{s,k} &= Z_{A,k} n_{0,k} = \bar n_k , \\
V_{A,k} &= - \frac{1}{2n_{0,k}} \frac{\partial^2 U_k}{\partial\mu^2}\biggl|_{n_{0,k}} , \\ 
Z_{C,k} &= - \frac{\partial^2 U_k}{\partial n\partial\mu}\biggl|_{n_{0,k}} = \lambda_k \frac{dn_{0,k}}{d\mu} ,
\end{split}
\label{ward1} 
\end{equation}
where $\bar n_k$ is the mean boson density and $n_{s,k}$ the superfluid density. Here we consider the effective potential $U_k$ as a function of the two independent variables $n$ and $\mu$. The first of equations~(\ref{ward1}) states that in a Galilean invariant superfluid at zero temperature, the superfluid density is given by the full density of the fluid~\cite{Gavoret64}. Equations~(\ref{ward1}) also imply that the Goldstone mode velocity $c_k$ coincides with the macroscopic sound velocity~\cite{Gavoret64,Pistolesi04,Dupuis09b}, \ie
\begin{equation}
\frac{d\bar n_k}{d\mu} = \frac{\bar n_k}{mc_k^2} .
\end{equation}
Since thermodynamic quantities, including the condensate ``compressibility''  $dn_{0,k}/d\mu$ should remain finite in the limit $k\to 0$, we deduce from (\ref{ward1}) that $Z_{C,k} \sim \lambda_k \sim k^{3-d}$ vanishes in the infrared limit, and 
\begin{equation}
\lim_{k\to 0} c_k = \lim_{k\to 0} \left( \frac{Z_{A,k}}{2mV_{A,k}} \right)^{1/2} .
\label{velir}
\end{equation}
Both $Z_{A,k}=\bar n_k/n_{0,k}$ and the macroscopic sound velocity $c_k$ being finite at $k=0$, $V_{A,k}$ (which vanishes in the Bogoliubov approximation) takes a non-zero value when $k\to 0$. 

\begin{figure}
\centerline{\includegraphics[width=6cm,clip]{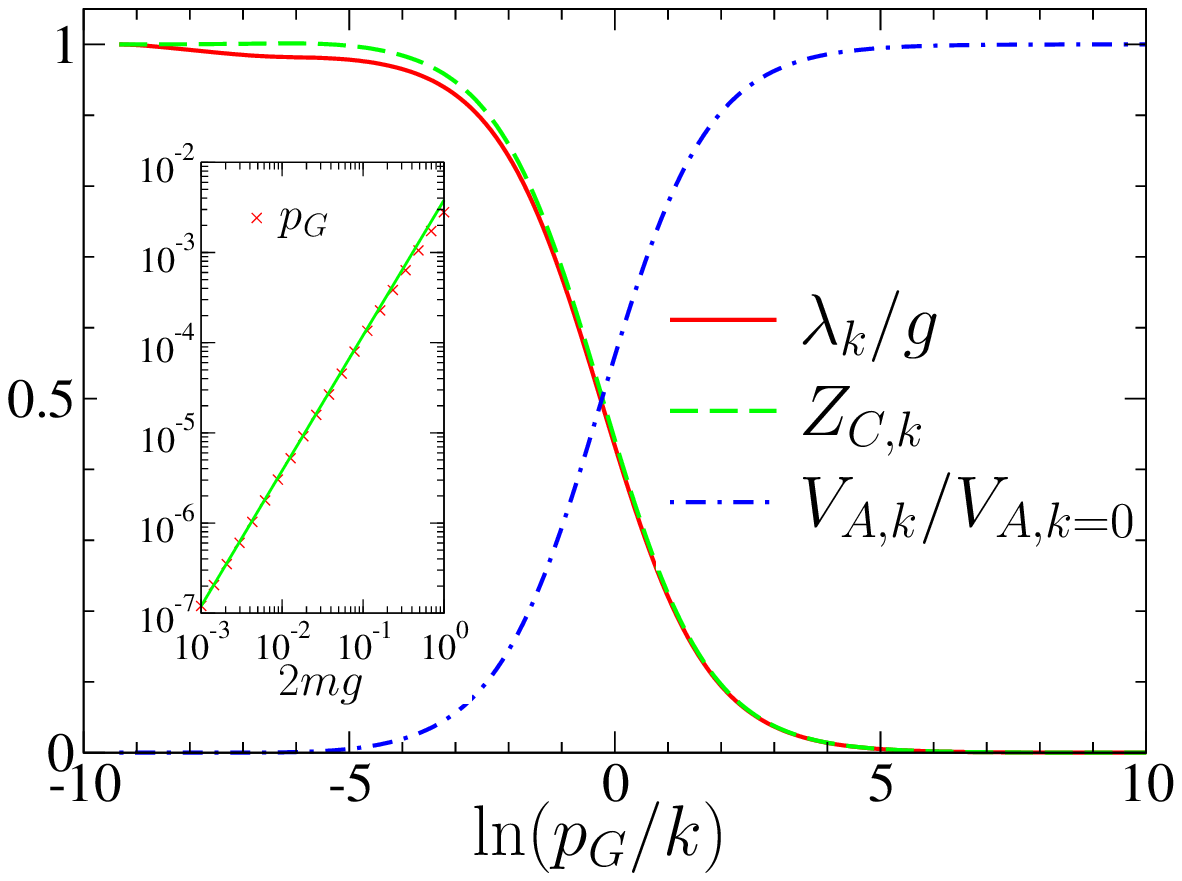}}
\caption{(Color online) $\lamb_k$, $Z_{C,k}$ and $V_{A,k}$ vs $\ln(p_G/k)$ where $p_G=\sqrt{(gm)^3\bar n}/4\pi$ for $\bar n=0.01$, $2mg=0.1$ and $d=2$ [$\ln(p_G/p_c)\simeq -5.87$]. The inset shows $p_G$ vs $2mg$ obtained from the criterion $V_{A,p_G}=V_{A,k=0}/2$ [the Green solid line is a fit to $p_G\sim (2mg)^{3/2}$].}
\label{fig_boson_flow_1} 
\centerline{\includegraphics[width=6cm,clip]{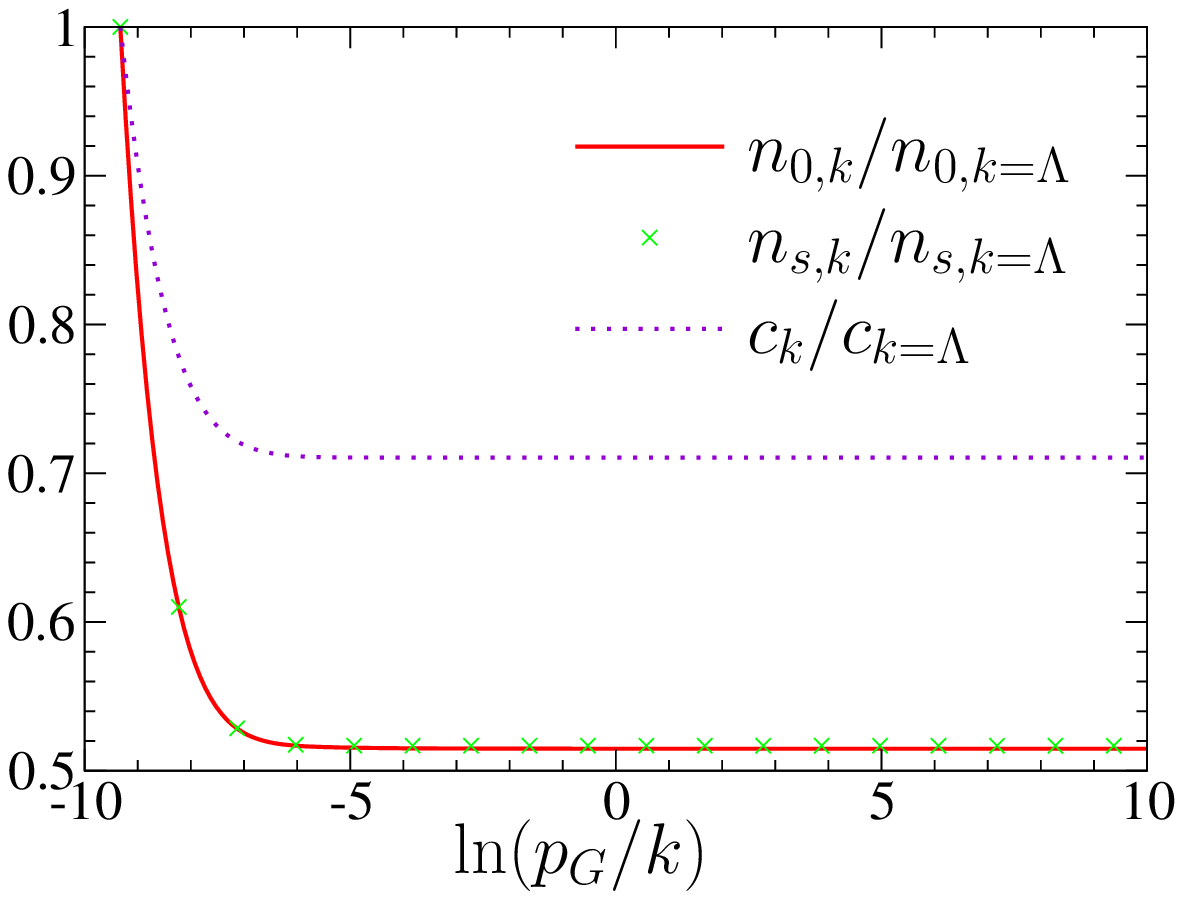}}
\caption{(Color online) Condensate density $n_{0,k}$, superfluid density $n_{s,k}$ and Goldstone mode velocity $c_k$ vs $\ln(p_G/k)$. The parameters are the same as in Fig.~\ref{fig_boson_flow_1}.} 
\label{fig_boson_flow_2} 
\end{figure}

The suppression of $Z_{C,k}$, together with a finite value of $V_{A,k=0}$ shows that the average effective action (\ref{effaction}) exhibits a ``relativistic'' invariance in the infrared limit and therefore becomes equivalent to that of the classical O(2) model in dimensions $d+1$. In the ordered phase, the coupling constant of this model vanishes as $\lamb_k\sim k^{4-(d+1)}$~\cite{Dupuis10}, which is nothing but our starting assumption (\ref{lambk1}). For $k\to 0$, the existence of a linear spectrum is due to the relativistic form of the average effective action (rather than a non-zero value of $\lambda_k n_{0,k}$ as in the Bogoliubov approximation)~\cite{note12}.

To obtain the $k=0$ limit of the propagators (at fixed $p$), one should in principle stop the flow when $k\sim \sqrt{\p^2+\w^2/c^2}$. Since thermodynamic quantities are not expected to flow in the infrared limit, they can be approximated by their $k=0$ values. As for the longitudinal correlation function, its value is obtained from the replacement $\lambda_k\to C(\w^2+c^2\p^2)^{(3-d)/2}$ (with $C$ a constant). From (\ref{de2}) and (\ref{ward1}), we then deduce the exact infrared behavior of the normal and anomalous propagators (at $k=0$),
\beq
\begin{split}
G_{\rm n}(p) ={}& - \frac{n_0mc^2}{\bar n} \frac{1}{\w^2+c^2\p^2} \\ &
- \frac{mc^2}{\bar n} \frac{dn_0}{d\mu} \frac{i\w}{\w^2+c^2\p^2} - \half G_{11}(p) , \\
G_{\rm an}(p) ={}& \frac{n_0mc^2}{\bar n} \frac{1}{\w^2+c^2\p^2} - \half G_{11}(p) , 
\end{split}
\label{de3}
\eeq
where
\begin{equation}
G_{11}(p) = \frac{1}{2n_0C (\w^2+c^2\p^2)^{(3-d)/2}} .
\label{de4}
\end{equation}
The hydrodynamic approach of Sec.~\ref{sec_hydro} correctly predicts the leading terms of (\ref{de3}) but approximates $dn_0/d\mu$ by $\bar n/mc^2$. On the other hand, it gives an explicit expression of the coefficient $C$ in the longitudinal correlation function (\ref{de4}).

\subsection{RG flows}
\label{subsec_bosons_num} 

The conclusions of the preceding section can be obtained more rigorously from the RG equation (\ref{rgeq}) satisfied by the average effective action. The flow of $\lamb_k$, $Z_{C,k}$ and $V_{A,k}$ is shown in Fig.~\ref{fig_boson_flow_1} for a two-dimensional system in the weak-coupling limit. We clearly see that the Bogoliubov approximation breaks down at a characteristic momentum scale $p_G\sim \sqrt{(gm)^3\bar n}$. In the Goldstone regime $k\ll p_G$, we find that both $\lamb_k$ and $Z_{C,k}$ vanish linearly with $k$ in agreement with the conclusions of Sec.~\ref{subsec_bosons_de}. Furthermore, $V_{A,k}$ takes a finite value in the limit $k\to 0$ in agreement with the limiting value (\ref{velir}) of the Goldstone mode velocity. Figure~\ref{fig_boson_flow_2} shows the behavior of the condensate density $n_{0,k}$, the superfluid density $n_{s,k}=Z_{A,k}n_{0,k}$ and the velocity $c_k$. Since $Z_{A,k=0}\simeq 1.004$, the mean boson density $\bar n_k=n_{s,k}$ is nearly equal to the condensate density $n_{0,k}$. Apart from a slight variation at the beginning of the flow, $n_{0,k}$, $n_{s,k}=Z_{A,k}n_{0,k}$ and $c_k$ do not change with $k$. In particular, they are not sensitive to the Ginzburg scale $p_G$. This result is quite remarkable for the Goldstone mode velocity $c_k$, whose expression (\ref{cdef}) involves the parameters $\lamb_k$, $Z_{C,k}$ and $V_{A,k}$, which all strongly vary when $k\sim p_G$. These findings are a nice illustration of the fact that the divergence of the longitudinal susceptibility does not affect local gauge invariant quantities~\cite{Pistolesi04,Dupuis09b}.

\section{The superfluid--Mott-insulator transition} 

\begin{figure}
\centerline{\includegraphics[width=6cm,clip]{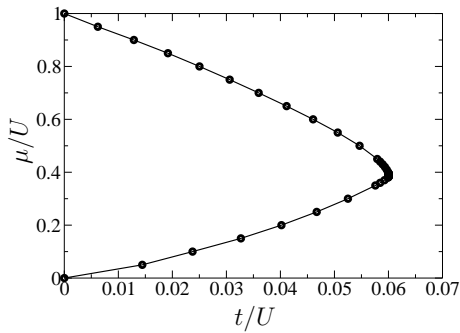}}
\caption{Phase diagram of the two-dimensional Bose-Hubbard model obtained from the NPRG~\cite{Rancon10} showing the first Mott lob corresponding to a density $\bar n=1$.}
\label{fig_bh} 
\end{figure}

The Bose-Hubbard model describes bosons on a $d$-dimensional lattice with Hamiltonian~\cite{Fisher89}
\begin{align}
\hat H ={}& -t \sum_{\mean{\r,\r'}} \left( \hat\psi^\dagger_\r \hat\psi_{\r'} + \hc \right) - \mu \sum_\r \hat n_\r \nonumber \\ & + \frac{U}{2} \sum_\r \hat n_\r (\hat n_\r - 1) , 
\end{align} 
where $\hat\psi_\r^{(\dagger)}$ is a creation/annihilation operator defined at the lattice site $\r$, $ \mean{\r,\r'}$ denotes nearest-neighbor sites, and $\hat n_\r=\hat\psi^\dagger_\r \hat\psi_\r$. This model can be studied within the NPRG framework using a formulation adapted to lattice models~\cite{Machado10,Rancon10}. Here we focus on the critical behavior at the superfluid--Mott-insulator transition. In the low-energy limit, we can take the continuum limit and express the average effective action as
\begin{align}
\Gamma_k[\phi^*,\phi] = \inttau\intr \bigl[ & \phi^* (Z_{C,k} \dtau - V_{A,k} \partial^2_\tau \nonumber \\ &  - Z_{A,k} t \nablabf^2 ) \phi + V_k(n) \bigr] , 
\end{align}
where $n(\r,\tau)=|\phi(\r,\tau)|^2$ and $\phi(\r,\tau)=\mean{\psi(\r,\tau)}$ denotes the superfluid order parameter. Near the transition, we can expand the effective potential $V_k(n)$ about $n=0$, 
\beq
V_k(n) = V_k(0) + \delta_k n + \frac{\lamb_k}{2} n^2 .
\eeq
The transition line in the $(t/U,\mu/U)$ plane is defined by $\delta\equiv\delta_{k=0}=0$ (Fig.~\ref{fig_bh}). The condensate density $n_0\equiv n_{0,k=0}$ is obtained from the minimum of $V(n)\equiv V_{k=0}(n)$. In the Mott phase ($\delta\geq 0$), $n_0$ vanishes while $n_0=-\delta/\lamb$ in the superfluid phase ($\delta\leq 0$). The mean boson density (\ie the mean number of bosons per site) is given by 
\begin{align}
\bar n &= - \frac{d}{d\mu} V(n_0) \nonumber \\ &=  \llbrace 
\begin{array}{lc} 
-\frac{dV(0)}{d\mu} & \mbox{(insulator)} , \\ 
-\frac{dV(0)}{d\mu} + \frac{\delta}{\lamb} \frac{d\delta}{d\mu} - \frac{\delta^2}{2\lamb^2} \frac{d\lamb}{d\mu} & \mbox{(superfluid)} .
\end{array}
\right.
\label{nbar} 
\end{align}
In the Mott phase, the density $-dV(0)/d\mu$ is pinned to an integer value. 

The existence of two universality classes at the quantum phase transition between the superfluid and the Mott insulator follows from symmetry arguments~\cite{Sachdev_book}. The last of the Ward identities (\ref{ward1}), which is associated to (local) gauge invariance~\cite{Dupuis09b}, can be expressed as 
\beq
Z_C = - \frac{d\delta}{d\mu} + \frac{\delta}{\lamb} \frac{d\lamb}{d\mu}
\eeq
in the superfluid phase. At the tip of the Mott lob where both $\delta$ and $d\delta/d\mu$ vanish, $Z_C$ vanishes. The effective action $\Gamma_{k=0}$ then exhibits a ``relativistic'' invariance and we expect the transition to be in the universality class of the $(d+1)$-dimensional $XY$ model with a dynamical exponent $z=1$. According to (\ref{nbar}), the lob tip ($\delta=d\delta/d\mu=0$) corresponds to a transition taking place at constant density. Away from the tip, the transition is accompanied by a change in density. $Z_C$ is finite and the transition is mean-field like for $d\geq 2$ with $z=2$ (with logarithmic corrections for $d=2$)~\cite{Sachdev_book}. 

\begin{table}
\renewcommand{\arraystretch}{1.5}
\begin{center}
\begin{tabular}{|c|c||c|c|c|c|c|}
\hline 
\multicolumn{2}{|c||}{} & $Z_{A,k}$ & $V_{A,k}$ & $Z_{C,k}$ & $\lamb_k$ & $n_{0,k}$
\\ \hline \hline 
\multicolumn{2}{|c||}{superfluid} & $Z_A^*$ & $V_A^*$ & $k$ & $k$ & $n_0^*$ 
\\ \hline 
critical & $XY$ & $k^{-\eta}$ & $k^{-\eta}$ & $k$ & $k^{1-2\eta}$ & $k^{1+\eta}$ 
\\ \cline{2-7} 
behavior & mean-field & $Z_A^*$ & $V_A^*$ & $Z_C^*$ & $|\ln k|^{-1}$ & $k^2|\ln k|^{-1}$ 
\\ \hline 
\multicolumn{2}{|c||}{insulator} & $Z_A^*$ & $V_A^*$ & $Z_C^*$ & $\lamb^*$ & 0 
\\ \hline 
\end{tabular}
\end{center}
\caption{Critical behavior at the superfluid--Mott-insulator transition and infrared behavior in the superfluid and Mott-insulator phases ($d=2$). The stared quantities indicate nonzero fixed-point values and $\eta$ denotes the anomalous dimension at the three-dimensional $XY$ critical point.} 
\label{table}
\end{table}

Table~\ref{table} summarizes the results obtained for the two-dimensional Bose-Hubbard model~\cite{Rancon10}. The NPRG provides a natural explanation for the critical behavior in the $XY$ universality class which is observed when the transition takes place at constant density (tip of the Mott lob). Indeed, since the infrared behavior in the superfluid phase is characterized by a relativistic symmetry, it is not surprising that this symmetry remains at the transition to the Mott insulator. By contrast, the Bogoliubov fixed point, which has a dynamical exponent $z=2$, is clearly a poor starting point to understand the superfluid--Mott-insulator transition at the lob tip. 

\section{Conclusion} 

Interacting bosons at zero temperature are characterized by two momentum scales: the healing (or hydrodynamic) scale $p_c$ and the Ginzburg scale $p_G$. For momenta $|\p|\ll p_c$, it is possible to use a hydrodynamic description in terms of density and phase variables. This description allows us to derive the correlation functions of the order parameter field without encountering infrared divergences. In the Goldstone regime $|\p|,|\w|/c\ll p_G$, density fluctuations play no role any more and both the transverse and longitudinal correlation functions are fully determined by transverse (phase) fluctuations. In this momentum and frequency range, the coupling between transverse and longitudinal fluctuations  leads to a divergence of the longitudinal susceptibility and singular self-energies. Nevertheless, in the weak-coupling limit, the Bogoliubov theory applies to a large part of the spectrum where the dispersion is linear ($|\p|\lesssim p_c$) and breaks down only at very small momenta $|\p|\lesssim p_G\ll p_c$. Moreover, thermodynamic quantities are not sensitive to the Ginzburg scale $p_G$ and can be deduced from the Bogoliubov approach.   

A direct computation of the order parameter correlation function (without relying on the hydrodynamic description) is possible, but one then has to solve the problem of infrared divergences which appear in perturbation theory when $|\p|\lesssim p_G$ and signal the breakdown of the Bogoliubov approximation. The NPRG provides a natural framework for such a calculation. It shows that in the Goldstone regime $|\p|,|\w|/c\ll p_G$, the system is described by an effective action with relativistic invariance similar to that of the $(d+1)$-dimensional classical O(2) model. This similarity sheds light on the critical behavior of the superfluid--Mott-insulator transition in the Bose-Hubbard model which belongs to the $XY$ universality class when the transition takes place at fixed density.


\begin{thebibliography}{40}
\bibitem[{not({\natexlab{a}})}]{note11}
\bibinfo{note}{Popov's original approach~\cite{Popov79} also takes into account
  the influence of the non-hydrodynamic modes ($|\p|\gtrsim p_c$) on the
  low-energy behavior of the normal and anomalous propagators. These modes
  however do not change the final results (\ref{Gn},\ref{Gan}) except for a
  minor modification in the expression of $G_{11}$ in three
  dimensions~\cite{Dupuis10}.}
\bibitem[{not({\natexlab{b}})}]{note12}
\bibinfo{note}{The emergence of the relativistic symmetry of the average
  effective action at low energy is connected to the singularities of the
  self-energies in the limit $p\to 0$~\cite{Dupuis10}. Thus we recover the fact
  that singular self-energies are crucial to obtain a linear spectrum in spite
  of the vanishing of the anomalous self-energies~\cite{Nepomnyashchii75}.}
\end{thebibliography}

\end{document}